\documentclass[11pt]{article}

\baselineskip
\usepackage{amssymb}
\usepackage{graphicx}
\usepackage{latexsym}
\usepackage{multirow}
\usepackage{epsfig}
\usepackage{amsmath}
\usepackage{color}
\usepackage[english]{babel}
\usepackage{comment}
\usepackage{caption}

\newcommand{\bra}{\left \langle}
\newcommand{\ket}{\right \rangle}

\newcommand{\eqa}{\begin{eqnarray}}
\newcommand{\ena}{\end{eqnarray}}

\newcommand{\Scl}{S_{\mbox{\tiny{cl}}}}

\newcommand{\eq}{\begin{equation}}
\newcommand{\en}{\end{equation}}

\newcommand{\Z}{\bf{Z}}

\newcommand{\SU}{\mathrm{SU}}

\newcommand{\de}{\partial}
\newcommand{\tmb}[1]{{\mbox{\tiny{#1}}}}
\newcommand{\ZZ}{\hbox{{\rm Z{\hbox to 3pt{\hss\rm Z}}}}}

\begin{document}

\begin{titlepage}
\begin{center}
{\Large\bf Ising string beyond Nambu-Goto}
\end{center}
\vskip1.3cm
\centerline{Alberto~Baffigo and Michele~Caselle}
\vskip1.5cm
\centerline{\sl Department of Physics, University of Turin \& INFN, Turin}
\centerline{\sl Via Pietro Giuria 1, I-10125 Turin, Italy}
\vskip0.5cm

\vskip1.0cm
\begin{abstract}

 A major result of the  Effective String Theory (EST) description of confinement is the so called "low energy universality," which states that the first few terms of the large distance expansion of any EST are universal and coincide with those of the Nambu-Goto action. Going beyond this  approximation is one of the most interesting open problems in the EST.
In the higher order terms beyond Nambu-Goto several important pieces of physical information are encoded, which could improve our understanding of the physical mechanisms behind confinement and of the physical degrees of freedoms which originate
the EST. 
In this paper we evaluate numerically the first two of these corrections  in the case of the three dimensional gauge Ising model. The first of them turns out to be negative: $\gamma_3=-0.00048(4)$, similar (but not equal) to the one recently measured in the $SU(2)$ Yang Mills theory in three dimensions and compatible with the bootstrap bound $\gamma_3 \geq -\frac{1}{768}$.

\end{abstract}

\end{titlepage}

\section{Introduction}
\label{sec:introduction}
One of the most promising approaches to understand and model the non-perturbative behaviour of confining Yang-Mills theories is the  "Effective String Theory" (EST) description in which the confining flux tube joining together a quark-antiquark pair is modeled as a thin vibrating string~\cite{Nambu:1974zg,Goto:1971ce,Luscher:1980ac,Luscher:1980fr,Polchinski:1991ax} .  

Recently, there has been a lot of progress in this context.  In particular it has been realized that the EST enjoys the so called "low energy universality" \cite{Meyer:2006qx,Luscher:2004ib, Aharony:2009gg, Aharony:2011gb, Dubovsky:2012sh,Billo:2012da, Gliozzi:2012cx, Aharony:2013ipa} which states that, due to the peculiar features of the string action and to the symmetry constraints imposed by the Poincar\'e invariance in the target space, the first few terms of its large distance  expansion are fixed and are thus universal.  This implies that the EST
is much more predictive than typical effective models and in fact its predictions have been be successfully compared in the past few years with lot of results on the interquark potential from Monte Carlo simulations in Lattice Gauge Theories (LGTs) (for recent reviews see for instance \cite{ Aharony:2013ipa,Brandt:2016xsp,Caselle:2021eir}).

At the same time it was recently realized that the simplest, Lorentz invariant, EST which is the well known 
Nambu-Goto model~\cite{Nambu:1974zg,Goto:1971ce} is an exactly integrable, irrelevant, perturbation of the bidimensional free Gaussian model \cite{Dubovsky:2012sh}, driven by the $T\bar T$ operator of the $D-2$ free bosons\footnote{We shall denote in the following with $D$ the number of spacetime dimensions of the target LGT and with $d\equiv D-1$ the number of spacelike directions.} \cite{Caselle:2013dra} which represent the transverse degree of freedom of the string. This observation stimulated lot of interesting results
even beyond the original application to Yang-Mills 
theories \cite{Dubovsky:2012wk, Dubovsky:2013gi, Dubovsky:2014fma, Cooper:2014noa, Dubovsky:2015zey, Conkey:2016qju, Dubovsky:2017cnj, Dubovsky:2018bmo}. In particular they are at the basis of a S-matrix bootstrap approach which can be used to constrain the EST action even beyond the Nambu-Goto approximation \cite{EliasMiro:2019kyf,EliasMiro:2021nul}.
  
Indeed, it is by now clear that the Nambu-Goto action should be considered only as a first order approximation of the actual EST describing the non-perturbative behavior of the Yang-Mills theories. Going beyond this approximation is one of the most interesting open problems in this context.
In the higher order terms beyond Nambu-Goto several important pieces of physical information are encoded. Their study could be of great importance to understand the physical mechanisms behind confinement or the physical degrees of freedoms which originate
the EST.

In particular, it is only by looking at these higher order terms that one may hope to find signatures, in the confining string, of the gauge group of the underlying LGT.
For this reason there is an increasing interest in exploring these corrections in different LGTs \cite{Caristo:2021tbk}. 
In this respect the three dimensional gauge Ising model that we shall study in this paper is a perfect laboratory to address this issue because,  thanks to the duality transformation with the the 3d Ising model, one can use innovative, powerful, algorithms to estimate these corrections. 
Moreover, its gauge symmetry is very different from standard $SU(N)$ gauge groups and allows to test, for instance, which is the effect of the discrete vs continuous 
gauge symmetries on the confining string. A final important reason is that the scaling function of the string tension, which will play a central role in the following, is known (thanks again to duality) with very high precision.

Due to the low energy universality these corrections appear at a very high order in the large distance expansion and their evaluation is a delicate task. One must reach very precise estimates of the ground state energy of the string for a wide range of distances and, possibly,  for a few different values of $\beta$ to test the correct scaling behaviour.  For these reasons we decided to evaluate them with two different approaches, using different algorithms (hierarchical Metropolis in one case and Swendsen-Wang in the other) choosing different observables (the ground state energy of the string in one case and its width in the other) and looking for an overall agreement of the final results between the two methods.

We were able to detect the first two corrections beyond Nambu-Goto which are described, as we shall see below, by the two parameters  $\gamma_3$ and $\gamma_5$ \cite{EliasMiro:2019kyf}.
The value of $\gamma_3=-0.00048(4)$ agrees with the S-matrix bound found in \cite{EliasMiro:2019kyf} and is similar, but slightly more negative than the value $\gamma_3|_{SU(2)}=-0.00037(6)$ found in 
\cite{Caristo:2021tbk} for the $SU(2)$ LGT in (2+1) dimensions.
The second correction is very small  $\gamma_5=3.0(4)\times 10^{-7}$ but not compatible with zero and its inclusion in the fit turned out to be mandatory to reach reasonable values of the reduced $\chi^2$.
{These values represent the first step toward a systematic study of $\gamma_3$ in  LGTs} 

 This paper is organized as follows. In the second section we shall define the problem and set the notations, then in the third section we shall recall a few basic results of the Effective String Theory of confinement. In the fourth section we shall present the 3d gauge Ising model and discuss its properties. Then, in sections five and six, we shall  discuss the two approaches that we used to evaluate $\gamma_3$ and $\gamma_5$ and present the main steps of the data analysis.
Finally the last section is devoted to a summary of the results and a few concluding remarks.

\section{General setting and notations}
\label{s2}

In the following we shall be mainly interested in finite temperature LGTs, which can be realized by 
{imposing periodic boundary conditions in the time direction for the bosonic field (and antiperiodic for fermionic ones)}. In a finite temperature setting the compactified "time" direction does not have any longer the meaning of time (recall that we are describing a system at equilibrium in the canonical ensemble) but its size $N_t$ is instead a measure of the inverse temperature of the system. Thus
 a lattice of size $N_s a$ in the spatial directions and $N_t a$ in the timelike direction represents the regularized version of 
a system of finite volume $V=(N_s a)^{d}$  at a finite temperature
$T=1/(N_t a)$. In the following we shall set the lattice spacing to $a=1$, and systematically neglect it.

In a finite temperature setting one can define
a new class of topologically non-trivial observables which are gauge invariant thanks  
to the periodic boundary conditions in the time direction: the
 Polyakov loops.
If we define the link dynamical variables of the gauge model as $U_\mu(\vec x,z)$ (where $\mu$ denotes the direction of the link and $(\vec x,z)$ its coordinates in the the lattice), we may define the Polyakov loop $P(\vec x)$ as follows:

\eq
P(\vec x)=\mbox{Tr}\prod_{z=1}^{N_t}U_t(\vec x,z)~~~.
\en

In a pure LGT  the Polyakov loop acquires a non-zero expectation value in the deconfined phase and is thus an order parameter of the finite temperature deconfinement transition.

 The value $\beta_c(N_t)$ of this deconfinement 
transition in a lattice of size $N_t=1/T$ can be used to define a new 
physical observable $T_c$ which is obtained by inverting $\beta_c(N_t)$. We obtain in this way, for each value of $\beta$, 
the lattice size in the time direction (which we shall call in
the following $N_{t,c}(\beta)$) at which the model undergoes the deconfinement
transition and from this the critical temperature $T_c(\beta)\equiv 1/N_{t,c}(\beta)$ as a function of $\beta$. We shall use this quantity  to set the scale of our simulations.
\vskip.3 cm

In the following we shall be mainly interested in correlators of Polyakov loops:
\eq
G(R,N_t)\equiv \langle P(x)P^\dagger(x+R) \rangle_{N_t},
\label{polya}
\en
where $R$ is the distance between the two Polyakov loops
 and the subscript $N_t$ in the expectation value  reminds that the correlator was evaluated on a lattice at  temperature $T=1/N_t$.
$G(R,N_t)$ can be used to define a finite temperature version of the interquark potential:
\eq
V(R,T)=-\frac{1}{N_t}\log{\langle P(x)P^\dagger(x+R) \rangle_{N_t}  }~~~~.
\label{potft}
\en

In the confining phase we expect, for large values of $R$, a linearly rising behaviour for $V(R,N_t)$,  which implies the following 
behaviour for the correlator:

\eq
\langle P(x)P^\dagger(x+R) \rangle_{N_t}  \sim {\rm e}^{-E_0(T) N_t R}
~~~.\label{area2}
\en
From  eq.(\ref{area2}) we may extract the ground state energy $E_0(T)$ of the confining flux tube joining the quark-antiquark pair. $E_0(T)$ depends on the temperature of the model and vanishes at the deconfinement transition.
The finite temperature behaviour  of $E_0(T)$ will play a major role in the rest of the paper.
It is interesting to notice that
the observable (\ref{polya}) has the topology of a cylinder whose circumference is fixed by the (inverse) temperature $N_t$ and the length by the interquark distance $R$.

\section{Effective String Theory}

\subsection{The Nambu-Goto action}

The behaviour of the flux tube in a confining LGT is well described by the Effective String Theory which models the flux tube as a thin vibrating string and allows to evaluate the contribution to the Polyakov loop correlator of the quantum fluctuations of this flux tube.

The simplest possible EST fulfilling the constraints imposed by the Lorentz invariance in the target space is 
the Nambu-Goto action \cite{Nambu:1974zg,Goto:1971ce} defined as follows:
\begin{align}
\label{NGaction}
S_\tmb{NG}= \sigma_0 \int_\Sigma d^2\xi \sqrt{g}~,
\end{align} 
where $~g\equiv \det g_{\alpha\beta}~$ and
\begin{align}
\label{NGaction2}
g_{\alpha\beta}=\partial_\alpha X_\mu~\partial_\beta X^\mu
\end{align} 

is the  metric induced on the reference world-sheet surface $\Sigma$ by the mapping $X_\mu(\xi)$ of the world sheet in the target space,
and $\xi\equiv(\xi^0,\xi^1)$
denote the worldsheet coordinates. This term has a simple geometric interpretation: it measures the area of the surface spanned by the string in the target space and has only one free parameter\footnote{We shall denote in the following the string tension with the index $0$ to avoid confusion with the spin variables of the Ising model.}:  the string tension $\sigma_0$.

In order to perform calculations with the Nambu-Goto action one has first to fix its reparametrization invariance.  The standard choice is the so called ``physical gauge''. In this gauge the two worldsheet coordinates are identified with the longitudinal degrees of freedom of the string: $\xi^0=X^0$, $\xi^1=X^1$, so that the string action can be expressed as a function only of the $(D-2)$ degrees of freedom corresponding to the transverse displacements, $X^i$, with $i=2, \dots , (D-1)$ which are assumed to be single-valued functions of the worldsheet coordinates. We shall comment below on the problems of this gauge fixing choice.

With this gauge choice  the determinant of the metric can be written as
\eqa
g&=&1+\de_0 X_i\de_0 X^i+\de_1 X_i\de_1 X^i\nonumber\\
&&\ \ \ +\de_0 X_i\de_0 X^i\de_1 X_j\de_1 X^j
-(\de_0 X_i\de_1 X^i)^2
\ena

and the Nambu-Goto action can then be written as a low-energy expansion in the number of derivatives of the transverse degrees of freedom of the string which, by a suitable redefinition of the fields, can be rephrased as a large distance expansion. 
 
\begin{equation}
S=\Scl+\frac{\sigma_0}{2}\int d^2\xi\left[\partial_\alpha X_i\cdot\partial^\alpha X^i + \cdots \right].
\label{action2NG}
\end{equation}

The first term of this expansion is exactly the gaussian action, i.e. a two dimensional
 Conformal Field Theory (CFT) of $D-2$ free bosons which represent the transverse degrees of freedom of the string, and the remaining terms combine themselves so as to give  an exactly integrable, irrelevant perturbation of this CFT \cite{Dubovsky:2012sh}, driven by the $T\bar T$ operator of the $D-2$ free bosons \cite{Caselle:2013dra}.

Thanks to this exact integrability, the partition function of the model can be written explicitely \cite{Luscher:2004ib, Billo:2005iv}. 
For the Polyakov loop correlator in which we are interested here \footnote{Similar expressions can be obtained also for the other relevant geometries: the Wilson loop \cite{Billo:2011fd} and the interface \cite{Billo:2006zg}}, the expression in $D$ space-time dimensions is, using the notations of  \cite{Luscher:2004ib, Billo:2005iv}:

\begin{equation}
\langle P(x)P^\dagger(x+R) \rangle_{N_t} 
  = \sigma_0^{\frac{4-D}{2}}\frac{N_t}{\pi}
   \sum_{n=0}^{\infty}w_n
  \left(\frac{{E}_n}{2R}\right)^{\frac{1}{2}(D-3)}
  K_{\frac{1}{2}(D-3)}({E}_nR)
\label{NG}
\end{equation}

where $K_{\frac{1}{2}(D-3)}$ is the modified Bessel function of order $\frac{D-3}{2}$, $R$ denotes, as above, the distance between the two Polyakov loops, $N_t$ the size of the lattice in the compactified direction, $w_n$ is the multiplicity of the closed string states which propagate from one Polyakov loop to the other, and $E_n$ their energies:
\begin{equation}
  {E}_n=\sigma_0 N_t
  \sqrt{1+\frac{8\pi}{\sigma_0 N_t^2}\left[-\frac{1}{24}\left(D-2\right)+n\right]}.
\label{energylevels}
\end{equation}

At large distances the correlator is dominated by the lowest state 
\eq
E_0=\sigma_0 N_t
  \sqrt{1-\frac{\pi (D-2)}{3\sigma_0 N_t^2}}
\label{E0}
\en  
whose multiplicity is $w_0=1$ and the Polyakov loop correlator, neglecting irrelevant constants, reduces to

 \begin{equation}
 \langle P(x)P^\dagger(x+R) \rangle_{N_t} 
  \sim
  \sigma_0^{\frac{4-D}{2}}{N_t}
  \left(\frac{E_0}{R}\right)^{\frac{1}{2}(D-3)}
  K_{\frac{1}{2}(D-3)}({E}_0R)
\label{NG2}
\end{equation}
which, in the $D=3$ case in which we are interested simplifies to

 \begin{equation}
 \langle P(x)P^\dagger(x+R) \rangle_{N_t} 
  \sim
  {N_t}\sqrt{\sigma_0}
  K_{0}({E}_0R)
\label{NG3}
\end{equation}

with
\eq
E_0=\sigma_0 N_t
  \sqrt{1-\frac{\pi}{3\sigma_0 N_t^2}}
\label{E0d3}
\en  

Thanks to the exponential term in the asymptotic expansion of the Bessel function, 
\eq
\label{K0_expansion}
K_0(z) = \sqrt{\frac{\pi}{2z}} e^{-z} \left[ 1 - \frac{1}{8z} + \frac{9}{128z^2}+ \mathcal{O}(z^{-3})  \right]
\en
we find at large distance, as expected, a linearly rising behaviour of the interquark potential controlled by the ground state energy $E_0$.
On top of this we have a set of subleading corrections, encoded in the asymptotic expansion of $K_{0}$, which represent a  specific, unique, signature of the Nambu-Goto action and must be taken into account when fitting the results of Monte Carlo simulations.

An important side consequence of this result is that we can extract from the tachyonic singularity of $E_0$ an estimate for the critical temperature $T_{c,NG}$  measured in units of the square root of the string tension $\sqrt{\sigma_0}$ \cite{Olesen:1985ej,Pisarski:1982cn} which is, for generic values of $D$,
\begin{equation}
\frac{T_{c,NG}}{\sqrt{\sigma_0}}=\sqrt{\frac{3}{\pi(D-2)}}
\end{equation}
and corresponds to the value of the ratio $\frac{T_{c,NG}}{\sqrt{\sigma_0}}$ for which the ground state energy $E_0$ vanishes.
Using this result we can rewrite the ground state energy as
\eq
E_0(T)|_{NG}=\frac{\sigma_0}{T}\sqrt{1-\frac{T^2}{T_{c,NG}^2}}
\label{eqtc}
\en
where we use the notation $|_{NG}$ to stress the fact that this is only the Nambu-Goto estimate for the ground state energy of the string, which we may expect to be modified by other terms in the EST action.

The estimate quoted above for the critical temperature turns out to be in remarkable (but not exact!) agreement with the results obtained from Monte Carlo simulations, both for non-abelian LGTs and for the three dimensional gauge Ising model. However, the remaining small deviations of $T_{c,NG}$ from the Monte Carlo results, together with the fact that the Nambu-Goto EST predicts (as can be seen looking at eq.(\ref{eqtc})) a deconfinement transition of the second order, with a mean field value for the critical index (which is in disagreement with the Monte Carlo results for all known LGTs) suggest that the Nambu-Goto action cannot be the end of the story and that some correcting terms beyond Nambu-Goto should be present in the EST. 

\subsection{Beyond Nambu-Goto}
\label{beyond}
It is by now clear that in the actual EST of the confining string the Nambu-Goto action is only the first term of an infinite series of contributions. Indeed, there are several reasons why the Nambu-Goto action is unsatisfactory and must be completed with some higher 
order correction.  Besides the above mentioned disagreement at the deconfinement transition, a major problem of the Nambu Goto action is that it is, so to speak,  too universal.  It predicts the same behaviour for all LGTs, with no dependence on the gauge group. Moreover, it is well known that the physical gauge fixing discussed above is anomalous and it is widely expected that this anomaly could be cured by higher order terms in the EST action.

The requirement of Poincarè invariance in the target space strongly constrains the terms which can be included in the EST beyond Nambu Goto~\cite{Meyer:2006qx,Luscher:2004ib, Aharony:2009gg, Aharony:2011gb, Dubovsky:2012sh,Billo:2012da, Gliozzi:2012cx, Aharony:2013ipa}.
In $D=3$ the first few allowed terms can be written as follows:

\eq
S_{EST}= \int_\Sigma d^2\xi \sqrt{g}~\left[\sigma_0 +  \gamma_1\mathcal{R} + \gamma_2 K^2 +  \gamma_3 K^4  \dots \right] \, 
\en
where the $\gamma_i$ are new coupling constants, 
 $\mathcal{R}$ denotes the Ricci scalar constructed from the induced metric, and $K$ is the extrinsic curvature, defined as $K=\Delta (g) X$, with  
\begin{equation}
\Delta(g)=\frac{1}{\sqrt{(g)}}\partial_a[\sqrt{(g)}g^{ab}\partial_b]
\end{equation}
the Laplacian in the space with metric $g_{\alpha\beta}$.
In principle the new coupling constants $\gamma_i$, should be fixed, as we do for $\sigma_0$, by comparing with experiments (or more likely, with computer simulations).

However this process is simplified by the observation that $K^2$ vanishes on shell and that the term proportional to $\cal{R}$ is a topological invariant in two dimensions. Since in the long-string limit in which we are interested one does not expect topology-changing fluctuations, both these terms can be neglected and the first non-trivial contribution appears only at higher orders~\cite{Aharony:2013ipa}.  This result is known as "low energy universality" \cite{Dubovsky:2012sh} and strongly constrains the form of the EST. It implies that the first correction beyond the Nambu-Goto action can only appear at the order $1/R^7$ (or $1/N_t^7$ in the finite temperature setting in which we are interested in this paper).
This explains why the Nambu-Goto model has been so successfull over these last forty years to describe the infrared behaviour of confining gauge theories despite its simplicity and why the deconfinement temperature predicted by Nambu-Goto is so close to the one obtained in Monte Carlo simulations. At the same time this explains why identifying these corrections is so difficult and only in the last few years it was possible to unambiguously detect them in Monte Carlo simulations \cite{Caristo:2021tbk,Athenodorou:2011rx,Dubovsky:2014fma,Chen:2018keo,Athenodorou:2016kpd}

The first non vanishing term $\gamma_3 K^4$ is only the first of an infinite sequence of terms, obtained combining higher order curvature invariants.
It turns out that the best way to organize these terms is to study the $2 \to 2$ scattering amplitude of the string 
excitations \cite{Dubovsky:2012sh}. 
It can be shown that in $D=3$ this S-matrix $S=e^{2i\delta}$ can be written in a particularly simple and elegant form:
\eq \label{SM1}
2\delta(s) = \frac{s}{4}   + \gamma_3 s^3 + 
\gamma_5 s^5+\gamma_7 s^7+i\gamma_8 s^8 +O(s^9)  \, , 
\en
where the the first term $s/4$ leads to the energy spectrum of the Nambu-Goto action,
the fact that the term proportional to $s^2$ is missing is the way in which the low energy universality is realized in this S-matrix approach
and the next non trivial term is exactly the S-matrix description of the $K^4$ correction mentioned above.

By using the analiticity properties of this S-matrix and requiring the UV completion of the underlying theory it is possible to obtain a set of important results on the ground state energy of the string\cite{Dubovsky:2012sh,EliasMiro:2019kyf}.

\begin{itemize}
\item $\gamma_8$ is not  a new independent parameter but it is proportional to $\gamma_3^2$
\item both the $1/N_t^7$ and the $1/N_t^9$ terms are controlled only by $\gamma_3$ and the next independent parameter $\gamma_5$ only appears
at the order $1/N_t^{11}$
\item It is possible to set bounds on these parameters.  In particular, defining
$$\tilde\gamma_n{=}\gamma_n+(-1)^{(n+1)/2}\frac{1}{n 2^{3n-1}}$$
one finds \cite{EliasMiro:2019kyf}

\eqa
\tilde  \gamma_3&\geq& 0\nonumber \\ 
\tilde \gamma_5&\geq& 4 \tilde \gamma_3^2 - \frac{1}{64}\tilde \gamma_3 \label{bound2}\\
\tilde \gamma_7&\geq &\frac{\tilde \gamma_5^2}{\tilde \gamma_3}+\frac{1}{4096}\tilde \gamma_3+\frac{1}{64}\tilde \gamma_5-\frac{1}{16}\tilde \gamma_3^2
\nonumber
\ena

which implies in particular
\eq
\gamma_3 \geq -\frac{1}{768}\,.
\label{bound1}
\en

\end{itemize}

From the S-matrix, by using the so called Thermodynamic Bethe Ansatz one can obtain \cite{EliasMiro:2019kyf} the following expression for the non universal corrections up to the order $1/N_t^{11}$

\eq
E_0(N_t)=\sigma_0N_t\sqrt{1-\frac{\pi}{3\sigma_0N_t^2}}-\frac{32 \pi ^6 \gamma _3}{225 \sigma_0^3N_t^7}
-\frac{64 \pi ^7 \gamma _3}{675 \sigma_0^4N_t^9}-\frac{\frac{2 \pi ^8 \gamma _3}{45} +\frac{32768 \pi ^{10} \gamma
   _5}{3969}}{\sigma_0^5N_t^{11}} 
\label{pred1}
\en

This is the expression which we shall compare with the results of our simulations.

\subsection{The boundary corrections problem and how to deal with it.}

It is clear from the previous section that measuring the $\gamma_i$ coeffcients on the lattice is a highly non-trivial task. In particular it is essentially impossible in the standard "zero temperature" scenario, in which the contribution of the effective string to the interquark potential manifests itself as an expansion in powers of $1/R$ and the corrections in which we are interested, which appear in this expansion at the order $1/R^7$, are shadowed by the boundary corrections which are proportional to $1/R^4$ \cite{Aharony:2010cx,Billo:2012da,Brandt:2010bw,Brandt:2017yzw,Brandt:2021kvt}.

There are in principle two ways to avoid this problem. 
\begin{itemize}
\item The first is to study observables without boundaries. This can be done for abelian gauge models using duality and looking at the finite size effects of the interface free energy (choosing interfaces with periodic boundary conditions)  
\cite{Caselle:1994df,Caselle:2007yc} . In the case of the Ising model this approach was recently discussed in \cite{Caselle:2016wsw} where it was shown that corrections beyond Nambu-Goto certainly exist in the gauge Ising model and are rather large. However it turned out to be difficult to quantify these corrections, most probably due to the interactions between nearby interfaces\footnote{The way in which these interfaces are generated in the (dual) gauge Ising model is by fixing antiperiodic boundary conditions in the transverse direction. This procedure generates an odd number of interfaces. This ensamble is usually studied assuming that they are far apart and do not interact, but when looking at very asymmetric geometries (which are needed in order to detect higher order corrections) the width of these interfaces grows linearly and negelecting interactions is most likely a too strong approximation.}.
 
\item The second is to study the model in the finite temperature regime (just below the deconfinement transition) in the limit of very large separation of the two Polyakov loops $(R>>N_t)$. It can be shown that in this regime the boundary  corrections become subleading and can be neglected \cite{Caselle:2011vk,Caselle:2021eir}. Moreover, this is exactly the limit discussed in the previous section, 
in which the results obtained with the S-matrix approach and TBA hold. 
Thus, by choosing this geometry in our simulations we shall be able to make immediate contact with eq. (\ref{pred1}), with no interference from the boundary terms and extract from the data reliable estimates for the $\gamma_i$ coefficents.
This was the approach recently used to study these corrections in the $SU(2)$ gauge model in three dimensions in \cite{Caristo:2021tbk}.
\end{itemize}

Once the geometry of the observables in which we are intersted is fixed, the remaining task is to obtain estimates for these observables precise enough to detect and quantify the tiny corrections in which we are interested. In this geometry this requires studying the system at large interquark distances and standard algorithms are affected by an exponentially decreasing signal to noise ratio in this limit.  The main advantage of studying abelian models is that, thanks to duality, it is possible to avoid this limitation and to study (using for instance non-local cluster algorithms as in \cite{Caselle:1996ii} or the so called "snake algorithm" \cite{deForcrand:2000fi}) Polyakov loops correlators at any interquark distance $R$ with the same signal to noise ratio. This is the main reason behind the choice of the Ising model as a laboratory to study the EST.

There is indeed a long track record of applications of this kind of methods to the 3d gauge Ising model to study subtle features of EST. In particular, in 
\cite{Caselle:1996ii,Caselle:2002ah,Caselle:2004jq,Caselle:2005xy} which may be considered as precursors of the present work, the $1/R^3$ correction to the interquark potential was precisely measured for the first time and shown to be exactly the one predicted by the Nambu-Goto action (in agreemeent with the low energy universality). Later the same approach was adopted for the 3d $U(1)$ model in \cite{Caselle:2014eka} 
and allowed to unambiguously detect corrections to the Nambu-Goto actions as the continuum limit was approached.

We shall discuss in sect.\ref{sect_snake} below the result of a study performed with the same snake algorithm used 
in \cite{Caselle:2002ah,Caselle:2004jq}.  This algorithm allowed us to obtain a first estimate of $\gamma_3$ which however turned out to be affected by a rather large statistical uncertainty.  In order to improve this uncertainty and to test the robustness of the result we decided to evaluate the same quantity with a completely different method, which was proposed a few years ago in \cite{Caselle:2010pf} and that we describe in detail in sect.\ref{sec:Zago}.

This second estimate agrees within the errors with the previous one, is more precise and allows to quantify with rather good precision even the next to leading correction $\gamma_5$. 

The agreement between the two estimates strongly supports the reliability of our analysis.  

In the following sections we shall first describe the main fetaures of the 3d gauge Ising model and then discuss the two approaches that we used to evaluate the $\gamma_i$ coefficients.

\section{The 3d Gauge Ising Model}
\label{sec:GaugeIsing}

The three dimensional Gauge Ising Model (also known as $\ZZ_2$ gauge model) was proposed over fifty years ago by Wegner~\cite{Wegner:1971app} as a tool for understanding the properties of lattice models with gauge symmetries. This model exhibits a local $\ZZ_2$ symmetry, which is realized by the choice of $\sigma_l \in \{1, -1\}$ as the dynamical $\ZZ_2$ link variables. The plaquette action, derived from the familiar Wilson action, is tailored specifically for the case of $\ZZ_2$ link variables and can be defined as follows: 

\eq
Z_{\mbox{\tiny{gauge}}}(\beta)=\sum_{\{\sigma_l=\pm1\}}\exp\left(-\beta S_{\bf{Z}_2}\right)
~.
\en
The action $S_{ \bf{Z}_2 }$ is a sum over all the plaquettes of  a cubic lattice,
\eq
S_{\bf{Z}_2 }=-\sum_{\Box}\sigma_\Box~~~,~~~
\sigma_\Box=\sigma_{l_1}\sigma_{l_2}\sigma_{l_3}\sigma_{l_4}~~.
\en

Despite its apparent simplicity the 3d Gauge Ising Model shares with more complex LGTs several important properties.
It is charaterized by a confining string with a non-trivial spectrum of string excitations \cite{Athenodorou:2022pmz,Caselle:2005vq}
and has a glueball spectrum very similar to the one of more complex 3d LGTs \cite{Agostini:1996xy} .  For these reasons it is a perfect laboratory to test, 
with high precision, non trivial properties of the confining strings in LGTs. 

This model is known to have a  bulk (i.e. at zero temperature)
deconfinement transition  at $\beta_c =0.76141330(6) $ (this value is obtained via duality from the critical temperature of the 3d Ising model quoted in \cite{Hasenbusch_2012},  see below).
For values of the coupling $\beta<\beta_c$ 
the model is in the confining phase, while for $\beta > \beta_c$ 
it is deconfined. The transition at $\beta=\beta_c$ 
is of  second order and, due to the duality relation (see below), it belongs to the same universality class of the standard magnetization transition of the 3d Ising model.
  This model also possesses an (infinite-order) 
``roughening transition'' at $\beta_r = 0.47542(1)$ \cite{Hasenbusch:1996eu}
(in the confined phase), which separates the strong coupling regime 
(for $\beta< \beta_r $) from the so-called ``rough phase'' 
(for $\beta_r < \beta < \beta_c $).

In the following we shall be interested in the behaviour of the model in the confining phase, in the scaling region near the critical point.
In particular we shall study the model at three different values of the
coupling constant $\beta$  (see tab.\ref{tab1}), for which the (finite) deconfinement temperature
 is known with high precision \cite{Caselle:1995wn} so as to be able to
precisely set the scale for the distances between  Polyakov loops and for the lattice
size in the time direction (i.e. the inverse of the finite temperature of the model).
These values are all located in the rough phase and close
enough to $\beta_c$ so as to be within the scaling
region.


The major reason of interest of this model is that it is related to the ordinary three dimensional Ising spin model by an exact duality 
mapping \emph{\`a la} Kramers and Wannier 
(see \cite{Savit:1979ny} for a general review on duality transformations):
\eq
\label{duality}
Z_{\mbox{\tiny{gauge}}} (\beta) \propto  Z_{\mbox{\tiny{spin}}} (\tilde{\beta}) \;\;\; , \;\;\; \mbox{with: $\tilde{\beta}=-\frac{1}{2}\log\left[\tanh(\beta)\right]$}
\en
where the Ising spin model is defined by the usual action:
\eq
\label{zspin}
Z_{\mbox{\tiny{spin}}} = \sum_{ \{ s_i \} } 
\exp \left( \tilde{\beta} \sum_{\bra i,j \ket} s_i s_j \right)
\en
where, as usual, the $s_i \in \{1,-1\}$ are $\ZZ_2$ spin variables, the $i$ 
and $j$ indices denote the sites of the dual lattice, the notation $\sum_{\bra i,j \ket}$ means that spin variables interact 
with their nearest-neighbours only and $\sum_{ \{ s_i \} }$ denotes the sum over spin configurations.

The critical temperature of the model is known with remarkable precision
$\beta_{c}= 0.22165462(2)$ \cite{Hasenbusch_2012}, moreover
thanks to the recent advances in the bootstrap approach \cite{Kos:2016ysd,Poland:2018epd} also the critical indices of the two relevant operators, the magnetization 
 $M$ and the energy $\epsilon$, are known with high precision:
$\Delta_{M} = 0.5181489 (10) $ and $\Delta_{\epsilon} = 1.412625 (10) $ respectively 
\cite{Kos:2016ysd,Poland:2018epd}. From $\Delta_{\epsilon}$  we can extract the critical index $\nu=\frac{1}{3-\Delta_{\epsilon}}=0.6299708...$.
Thanks to duality this is the same critical index which drives the critical behaviour of the string tension in the 3d gauge Ising model.  More precisely $\sigma(\beta)\sim \sigma_c(\beta_c-\beta)^{2\nu}$.  We shall further discuss the scaling behaviour of the string tension in the appendix.

The main reason of interest of this mapping is that a similar construction can be performed also 
in presence of  external source terms for the gauge model
(for instance, a pair of Polyakov loops). This can be easily realized by 
introducing sets of topological defects in the spin system. As a result, 
it is possible to show \cite{Caselle:2002ah} that, for instance,  the Polyakov loop correlator in which we are interested 
is mapped into the partition function of the spin system with anti-ferromagnetic coupling on 
a well defined set of links:
\eq
\label{Gspin}
G(r)\equiv \langle P(x)P^\dagger(x+R) \rangle_{N_t} = \frac{Z_{\mbox{\tiny{spin}},Q\bar{Q}}(R,N_t)}{Z_{\mbox{\tiny{spin}}}}
\en
with:
\eq
\label{zspindefects}
Z_{\mbox{\tiny{spin}},Q\bar{Q}}(R,N_t) = \sum_{ \{ s_i \} } 
\exp \left( \tilde{\beta} \sum_{\bra i,j \ket} J_{\bra i,j \ket} s_i s_j \right)
\en
where the value of the $J_{\bra i,j \ket}$ 
coupling is $+1$ everywhere, except on a set of bonds, which pierce a surface 
(in the direct lattice) having the source worldlines as its boundary: 
for such a set of bonds, $J_{\bra i,j \ket}=-1$.

Similar mappings can be constructed essentially for any observable of interest in the gauge model, from  Wilson loops 
to glueball correlators\footnote{ For instance it can be shown in this way that
the glueballs of the Ising gauge model are mapped into bound states of the fundamental scalar in the spin model 
\cite{Caselle:2001im}.}.

Both the numerical algorithms that we shall use in the following 
exploit this duality of the model, simulating the Ising 
spin system, and measuring (the first algorithm) ratios of partition functions associated with 
different stacks of defects (which we shall use to express the expectation 
values of Polyakov loops pairs in the original gauge model) or expectation values of spins in presence of the defect surface (the second algorithm). 

A particularly useful advantage of numerical simulations in the dual setting 
is the fact that this method overcomes the problem of exponential 
signal-to-noise ratio decay, which is usually found when studying the 
interquark potential $V(r)$ at larger and larger distances.

Another important feature is that, thanks to duality, the valus of the Polyakov loop correlators that we obtain in this way are not affected by  corrections due to the periodic boundary conditions in the spacelike directions.  This greatly simplifies the study of these correlators (no additional terms must be included in the fits to keep into account these corrections) and allow to study the system for 
(relatively) small lattice size in the spacelike directions.

We perfomed simulations for the three values of $\beta$ quoted in tab.\ref{tab1}: $\beta=0.751800,~0.756427,~0.758266$
which were chosen because for these values the deconfinement temperature is known with very high precision and
coincides with $1/T_c=L_c=8,12,16$ respectively~\cite{Caselle:1995wn}. 

\begin{table}[ht]
\centering %
\begin{tabular}{| c | c | c | c | c |} 
\hline
$\beta$ & $\tilde{\beta}$ & $N_{t,c}$ & $\sigma$ & $\alpha$\\
\hline
 0.751800  & 0.226104 & 8 &  0.0105255(11) & 0.4576(4)\\
\hline
 0.756427  & 0.223951 & 12 & 0.0046384(26) & 0.3887(3)\\
\hline
 0.758266  & 0.223101 & 16 & 0.0026043(53) & 0.3464(2)\\
\hline
\end{tabular}
\caption{Some information on the three values of $\beta$, listed in the first column, which we chose for the simulations. In the second column we report the corresponding values of $\tilde\beta$ for the (dual) 3d Ising model. In the last three colums columns we  report respectively the (inverse of) the deconfinement temperature, the string tension (taken from ref.\cite{Caselle:2007yc}) and the value of $\alpha$ (see below for the definition of $\alpha(\beta)$ and its evaluation from the scaling function of the model).}
\label{tab1}
\end{table}

\section{Analysis with the snake algorithm}
\label{sect_snake}

The first method that we used to estimate $\gamma_3$ is the Ising implementation of the snake algorithm \cite{deForcrand:2000fi} discussed in detail in \cite{Caselle:2002ah}. The main feature of the algorithm is the hierarchical organization of the updates.  In our particular implementation we chose five sublattice levels with size  $\{6, 13, 17, 21, 24\}$ lattice spacings respectively.

We studied the first two values of $\beta$ reported in tab.\ref{tab1} corresponding to a deconfinement temperature of $N_{t,c}=8$ and $N_{t,c}=12$ respectively.  Details on the simulations are reported in tab.\ref{tabB1}.
For each $\beta$  we studied seven values of $N_t$ just above the critical value $N_{t,c}$ (i.e. just below the deconfinement temperature). Then for each value of $N_t$ we simulated 8 different values of $R$ as reported in the table. The value of the lattice size in the space direction was chosen to be ten times the value of $N_{t,c}$ to avoid finite size effects (which in any case are strongly suppressed thanks to the duality transformation). The values of $R$ were chosen so as to make higher order energy levels in eq.(\ref{NG}) negligible within the errors. Thus we could fit the $R$ dependence of our Polyakov loop correlators using eq.(\ref{NG3}). From the snake algorithm we directly obtain the ratio of two nearby correlators $F(R,N_t)=G(R+1,N_t)/G(R,N_t)$ which we thus fitted with the following expression:
\eq
F(R,N_t)=\frac{K_0(E_0(N_t) (R+1))}{K_0(E_0(N_t) R)}
\label{Bfit1}
\en
with $E_0(N_t)$ as the only free parameter of the fit.
For all values of $N_t$ we found good values of the reduced $\chi_2$.  
We report in tab. \ref{Btab2}, as an example of the results obtained with the snake algorithm, the values obtained for  the largest Polyakov loops correlators that we studied i.e. those for $\beta=0.756427$ and $N_t=24$.  As anticipated there is no increase in the signal to noise ratio as $R$ increases and we could estimate the ratio of the two Polyakov loop correlators with less than 1\% error for areas as large as $24\times 84$ lattice spacings.

\begin{table}[ht]
\centering %
\begin{tabular}{| c | c | c | c | c | c |} 
\hline
$\beta$ & $N_{t,c}$ & $N_t$ & $R$ & $N_s$  \\
\hline
 $0.751800$  & 8 &  9,10,11,12,14,16,18 & 8, 12, 16, 20, 24, 32, 40, 48 & 80 \\
\hline
 $0.756427 $ & 12 &  13,14,15,16,18,20,24 & 18, 24, 30, 36, 48, 60, 72, 84 & 120 \\
\hline
\end{tabular}
\caption{Some information on the simulations.}
\label{tabB1}
\end{table}

 \begin{table}[ht]
  \centering
\begin{tabular}{ | c | c | }
                            \hline
    $R$ &   $\frac{G(R+1)}{G(R)}$  \\  \hline
18 &  0.8914(29)  \\ \hline
24 &  0.8969(31) \\ \hline
30 &  0.9003(32) \\ \hline
36 &  0.9030(33)  \\ \hline
48 &  0.9036(33)  \\ \hline
60 &  0.9035(33)  \\ \hline
72 &  0.9070(35)  \\ \hline
84 &  0.9085(35)  \\ \hline
\end{tabular}
\caption{Results of the snake algorithm for $\beta=0.756427$ and $N_t=24$.}
  \label{Btab2}
    \end{table}

We report in tab.\ref{Btab3} and \ref{Btab4} the results of these fits.

\begin{table}[h!]
  \centering
\begin{tabular}{ | c | c | c | }
\hline 
 $N_{t}$ & $E_{0}$   & $\chi^{2}_r$ \\  \hline
9 & 0.0226(11)&  0.90\\  \hline
10 & 0.0419(10) & 0.13\\  \hline
11 & 0.0587(9) & 0.33\\  \hline
12 & 0.0776(9)  & 0.34\\  \hline
14 & 0.1058(9)  & 0.16\\  \hline
16 & 0.1337(10)  & 0.08\\  \hline
18 & 0.1596(10)  & 0.05\\  \hline
\end{tabular}
\caption{Results of the fit with eq.(\ref{Bfit1}) for $\beta=0.751800$. In the last column the reduced $\chi^2$ of the fits.}
\label{Btab3}
    \end{table}

  \begin{table}[h!]
  \centering
\begin{tabular}{ | c | c | c | }
\hline 
 $N_{t}$ & $E_{0}$   & $\chi^{2}_r$  \\ \hline
13 & 0.0099(10) & 0.55 \\ \hline
14 & 0.0200(10) & 0.44 \\ \hline
15 & 0.0263(10) & 0.26 \\ \hline
16 & 0.0370(11) & 0.31 \\ \hline
18 & 0.0494(10) & 0.29 \\ \hline
20 & 0.0641(12) & 0.21 \\ \hline
24 & 0.0906(12) & 0.14 \\ \hline
\end{tabular}
\caption{Results of the fit with eq.(\ref{Bfit1}) for $\beta=0.756427$. In the last column the reduced $\chi^2$ of the fits.}
 \label{Btab4}  
  \end{table}

These are the values that we compared with the expectation of  eq.(\ref{pred1}).  
 We performed two types of fit. In the first we kept as free parameters only $\sigma_0$ and $\gamma_3$. Accordingly we truncated the square root of the Nambu-Goto  
action to the same order $O(N_t^{9})$ at which the $\gamma_3$ terms appears. Results of these fits are reported in tab.\ref{Btab5}
In the second we included also $\gamma_5$ and, accordingly, truncated the square root to the order $O(N_t^{11})$.
Results of these fits are reported in tab.\ref{Btab6}.

\begin{table}[ht]
\centering %
\begin{tabular}{| c | c | c | c | }
\hline
$\beta$ & $\sigma_0$ & $\gamma_3$ & $\chi^2_r$  \\
\hline
 0.751800  & 0.010671(30) & -0.000357(13) & 1.81 \\
\hline
 0.756427  & 0.004771(27) & -0.000329(13) & 1.62 \\
\hline
\end{tabular}
\caption{Results of the fits of $E_0(N_t)$ according to eq.(\ref{pred1}) truncated at the order $O(N_t^{9})$ .}
\label{Btab5}
\end{table}

\begin{table}[ht]
\centering %
\begin{tabular}{| c | c | c | c | c |} 
\hline
$\beta$ & $\sigma_0$ & $\gamma_3$ & $\gamma_5$ & $\chi^2_r$ \\
\hline
 0.751800  & 0.010629(35)  & -0.00061(11) & 0.00000042(11) & 1.00 \\
\hline
 0.756427  & 0.004740(34) & -0.00051(13) & 0.00000032(12) & 1.53 \\
\hline
\end{tabular}
\caption{Results of the fits of $E_0(N_t)$ according to eq.(\ref{pred1}) truncated at the order $O(N_t^{11})$ .}
\label{Btab6}
\end{table}

A few observations on this result:
\begin{itemize}

\item In the first type of fits the values of $\sigma_0$ that we obtain are definitely larger (more than three standard deviations) than the expected ones. Accordingly, if we try to fit the data keeping $\sigma_0$ to the expected value we also found very large values of $\chi^2$. Moreover, the fact that the reduced $\chi^2$ is larger than 1 suggests that the inclusion of the next order term in the fits could lead to non negligible corrections and in fact the values of $\gamma_5$ in the fits truncated at  $O(N_t^{11})$ are different from zero within the errors.
We also tried to fit the data truncating the expansion at the order $O(N_t^{7})$ and keeping only the first correction proportional to $\gamma_3$, but we found values of $\sigma_0$ even further away from the expected values.

\item Including the $O(N_t^{11})$ term in the fit the values of $\sigma_0$ that we obtain move toward the expected values and for both values of $\beta$ are within two standard deviations from the values reported in tab.\ref{tab1}\footnote{This trend suggests that including even the $\gamma_7$ correction we could reach the correct values of $\sigma_0$. We tried to include also this correction in the fits, but the value of $\gamma_7$ turned always to be compatible with zero within the errors.}.

\item 
Accordingly, in all these fits, if we force $\sigma$ to the known values we always find unacceptably high values of the $\chi^2$ which were associated to large deviations in the best fit values of $\gamma_3$ and $\gamma_5$.

\item The values of $\gamma_3$ and $\gamma_5$ that we find for the two values of $\beta$ are compatible with each other within the errors.

\end{itemize}

Looking at the results we see that the estimates of $\gamma_3$ and $\gamma_5$ are affected by rather large errors and are strongly influenced by the value of $\sigma$. This is due to the fact that the fit is dominated by the Nambu-Goto part of the fitting function and in particular by the $\sigma N_t$ term and by the L\"uscher correction. It seems difficult to improve the overall precision of the result in the framework discussed in this section since the simulations (due to the peculiar structure of the snake algorithm and the need to increase the size of the lattice in the inverse temperature direction) become more and more expensive as $\beta$ moves toward the continuum limit. For this reasons we decided to approach the task following a different strategy which could partially overcome these problems.

\section{Using the mean flux density in presence of the Polyakov loops to estimate the EST ground state energy.}
\label{sec:Zago}

To avoid the above problems we tried a completely different approach.  Following \cite{Caselle:2010pf}
instead of looking at
 the interquark potential, we studied the changes induced in 
the flux configuration by the presence of the Polyakov loops. 
We shall show below that as a consequence of this choice the explicit dependence on $\sigma_0$ and on the L\"uscher term vanish. This makes this observable an
unique tool to explore higher order corrections. 

Another reason of interest of 
this approach is that it is deeply related to another important issue of the effective string description of LGTs, i.e.
the study of the flux tube thickness. 
It can be shown that in the high temperature regime in which we are
presently interested the width of the flux tube increases linearly with the interquark distance and not logarithmicaly as one would
naively expect~\cite{Gliozzi:2010zv,Allais:2008bk,Caselle:2010zs}. This linear increase is  related to the linear increase in the flux energy that we observe 
here. In both cases the slope is temperature dependent and contains information on the higher order effective string
corrections in which we are interested.

The lattice operator which measures the flux through a plaquette $p$ in presence of two Polyakov loops $P$, $P'$ for a generic LGT is:
\eq 
\bra\phi(p;P,P')\ket=\frac{\bra P P'^\dagger~U_p\ket}{\bra PP'^\dagger \ket}-\bra U_p\ket 
\label{flux2} 
\en 
where $U_p$ is the trace of the ordered product of the link variables along the plaquette and in our case coincides with the product 
$\sigma_\Box$ introduced above. This is the quantity which is typically used to study the profile of the flux tube. 
In our analysis we are actually interested in a much simpler observable, the mean flux density, i.e. the sum of $\phi(p;P,P')$ over all the positions and orientations of the plaquettes, normalized to
the number of plaquettes of the lattice. 
Due to translational invariance this quantity will depend only on the distance $R$ between the two Polyakov loops and on the inverse temperature $N_t$. 
Let us define
\eq
\bra\Phi(R,N_t)\ket=\frac{1}{N_p}\sum_p \frac{\bra P P'^\dagger~\sigma_\Box\ket}{\bra PP'^\dagger \ket} - \bra \sigma_\Box\ket
\en
where $N_p=3N_s^2L$ denotes the number of plaquettes of the lattice.

It is easy to see from the definition of $G(R,N_t)$:
\eq
G(R,N_t)=\langle P^\dagger (R) P(0) \rangle_{N_t} = \frac{1}{Z}\sum_{conf} P^\dagger (R) P(0) ~ e^{\beta \sum_p \sigma_\Box}  
\label{neq1}
\en
that the mean flux density
$\bra\Phi(R,N_t)\ket$ can be written as:
\eq
\bra\Phi(R,N_t)\ket=\frac{1}{N_p}\frac{d}{d \beta} \log G(R,N_t)~~~.
\label{defPhi}
\en
Since $\beta$ appears in the  observable only through the scale $\sigma_0(\beta)$ the above equation can be rewritten as
\eq
\bra\Phi(R,N_t)\ket=\frac{1}{N_p}\frac{d\sigma_0}{d \beta} \frac{d}{d \sigma_0}\log G(R,N_t)~~~.
\label{defPhi2}
\en

This choice has two important consequences:
\begin{itemize}
\item
In the term proportional to $R$ the string tension,  which was the major source of systematic uncertainty in the previous calculation, is substituted by its derivative with respect to $\beta$, which can be evaluated with high confidence thanks to the very precise knowledge we have of the scaling behaviour of $\sigma_0(\beta)$ (see below).
\item
The first string correction (the "L\"uscher term") which is universal and does not depend on $\sigma_0$ disappears.
\end{itemize}

This makes the above observable a perfect tool to explore higher order corrections of the effective string.

From eq.(\ref{NG3}) we have

\eq
N_s^2\bra\Phi(R,N_t)\ket=\alpha(\beta) \left(\frac{1}{2N_t\sigma_0}+\frac{K'_0(E_0R)}{K_0(E_0R)} \frac{R}{N_t} \frac{d E_0}{d \sigma_0}\right)~~~.
\label{defPhi3}
\en

where $K'_0$ denotes the derivative of the $K_0$ Bessel function, and $\alpha$ is defined as
\eq
\alpha(\beta)=-\frac{1}{3}\frac{d \sigma_0}{d \beta} ~~~.
\label{defalpha}
\en
Using the identity $K'_0(z)=-K_1(z)$ the logarithmic derivative $K'_0(z)/K_0(z)$ can be expanded in powers of $1/z$  as follows:
\eq
\frac{K'(z)}{K(z)}=-(1+\frac{1}{2z}-\frac{1}{z^2}+\cdots)
\en
which gives:

\eq
N_s^2\bra\Phi(R,N_t)\ket=\alpha(\beta)\left(R A(N_t) + B(N_t) + C(N_t)/R\right)~~~,
\label{defPhi4}
\en
where the three functions $A,B$ and $C$ are defined as follows
\eq
A(N_t)=\frac{1}{N_t} \frac{d E_0}{d \sigma_0}
\en
\eq
B(N_t)=\frac{1}{2N_t E_0}\frac{d E_0}{d \sigma_0} - \frac{1}{2N_t\sigma_0}
\en
\eq
C(N_t)=-\frac{1}{8N_t E^2_0}\frac{d E_0}{d \sigma_0} 
\en

A crucial role in the analysis is played by $\alpha(\beta)$,  a precise estimate of this quantity allows to strongly constrain the results of the fits.
$\alpha$ can be extracted from the scaling function of the model and in its determination we leverage the very precise knowledge we have of this scaling function, thanks to the bootstrap results for the critical indices of the 3d Ising model. A detailed derivation can be found in \cite{Caselle:2010pf}.  We report for completeness the main steps of the derivation in the appendix \ref{AppendixA}. The values we used in the fit are listed in tab. \ref{tab1}.
Once the value of $\alpha$ is fixed,
we can use the values we obtain for $A(N_t)$ to estimate the corrections in which we are interested\footnote{In principle we could use also $B(N_t)$ or $C(N_t)$ to extract this information,  but these terms, due to their $R$ dependence may be affected by boundary corrections,  moreover their estimates from the simulations are much less precise than those of $A(N_t)$, so we neglected them in the following.}.  
By setting $x=\frac{\pi}{3\sigma_0 N_t^2}$
we see that we can express the Nambu-Goto expectation for $A$ (see eq.(\ref{E0d3})) as
\eq
A(N_t)_{NG}=\frac{1-\frac{x}{2}}{\sqrt{1-x}}= 1+ \frac{x^2}{8}+ \frac{x^3}{8}+ \frac{15x^4}{128}+ \frac{7x^5}{64}+ \frac{105x^6}{1024}+\cdots
\en
As anticipated the expansion starts at the order $x^{-2}$ i.e. $N_t^{-4}$, this makes this observable particularly suited to evaluate higher order corrections.

If we introduce the corrections beyond Nambu-Goto (see eq.(\ref{pred1})) we find
\eq
A(N_t)=A(N_t)_{NG}+ \frac{864\pi^2}{25}\gamma_3 x^4+\frac{2304\pi^2}{25}\gamma_3 x^5 + (162 \pi^2 \gamma_3+ \frac{1474560\pi^4}{49}\gamma_5 )x^6
\label{zfit1}
\en

where the expansion of $A(N_t)_{NG}$ is truncated at the same order at which the additional terms proportional to $\gamma_3$ and $\gamma_5$ appear.

This is the function which we shall use to fit the results of our simulations.

\subsection{Numerical simulations}

To estimate the function $A(N_t)$ we used again duality and mapped the Polyakov loops correlator
into the partition function of a 3d Ising spin model in which we changed the sign of the coupling 
of all the links dual to the surface bordered by the two Polyakov loops.
This is the same method which was used in  \cite{Caselle:1995fh,Allais:2008bk} to estimate the width of the flux tube.

We then estimated $\bra\Phi(R,N_t)\ket$ by simply evaluating the mean value of the plaquette in presence of these frustrated links. 
We chose periodic boundary conditions in the original 
gauge Ising model. These b.c. are mapped by duality into a mixture of periodic and antiperiodic b.c. in the dual spin model. However we always chose $N_s$ large enough
to eliminate any contribution from the antiperiodic sectors (which are depressed by terms proportional to $e^{-\sigma_0 N_s N_t}$ or $e^{-\sigma_0 N_s^2}$). 

Since, as discussed above, we are interested in the following only in the
term proportional to $R$ in $\bra\Phi(R,N_t)\ket$, we may neglect the 
disconnected component $\bra U_p\ket$  in the evaluation of  $\bra\Phi(R,N_t)\ket$.

Details on the simulations can be found in tab.\ref{ztab1}

We perfomed simulations for all the three values of $\beta$ reported in tab.\ref{tab1}. For each value of $\beta$ and $N_t$ we simulated several values of the distance $R$ between the two Polyakov loops. 

For each simulation we used $10^5$ iterations to thermalize the lattice and then performed $10^6$ measures using a Swendsen-Wang algorithm.
The values of $R$ were chosen large enough so as to make the last term in eq.(\ref{defPhi4}) negligible, thus allowing to perform a simple linear fit to extract the values of $A(N_t)$. We report in tab.\ref{ztab2} an example of the data we obtained from the simulations (to allow a comparison, we chose the same values of $\beta$ and $N_t$ reported in tab.\ref{Btab2}) and in tab.s \ref{ztab3a},\ref{ztab3b} and \ref{ztab3c} the values of $A(N_t)$ extracted from these linear fits.

We then fitted these values with eq.(\ref{zfit1}) keeping as only free parameters $\gamma_3$ and $\gamma_5$. Results are reported in tab.\ref{ztab4}.
Thanks to the high precision in the determination of $\alpha$, the systematic uncertainty on $\gamma_3$ and $\gamma_5$ due to the uncertainty on $\alpha$ is negligible and we quote in tab.\ref{ztab4} only the statistical errors of the fits.

\begin{table}[ht]
\centering %
\begin{tabular}{| c | c | c | c | c | c |} 
\hline
$\beta$ & $N_{t,c}$ & $N_t$ & $R$ & $N_s$  \\
\hline
 $0.751800$  & 8 &  9,10,11,12,16,20,24 &  16, 24, 32, 40, 48, 64 & 128 \\
\hline
 $0.756427 $ & 12 &  13,14,15,16,18,20,24 & 36, 48, 60, 72, 84, 96 & 192 \\
 \hline
 $0.758266 $ & 16 &  17,18,19,20,21,22,24,32,48 & 32, 48, 64, 80, 96, 112 & 256 \\

\hline
\end{tabular}
\caption{Some information on the simulations.}
\label{ztab1}
\end{table}

 \begin{table}[ht]
  \centering
\begin{tabular}{ | c | c | }
                            \hline
    $R$ &  $\bra\Phi(R,N_t)\ket$   \\  \hline
36 &  0.927589(9)  \\ \hline
48 &  0.927717(9)   \\ \hline
60 &   0.927872(9)  \\ \hline
72 &   0.927978(9)   \\ \hline
84 &   0.928115(9) \\ \hline
96 &   0.928247(9)   \\ \hline
\end{tabular}
\caption{Results of the  algorithm for $\beta=0.756427$ and $N_t=24$.}
  \label{ztab2}
    \end{table}

  \begin{table}[h!]
  \centering
\begin{tabular}{ | c | c | }
\hline 
 $N_{t}$ & $A(N_t)$  \\ \hline
 9 & 1.088(23)  \\ \hline
 10 & 1.071(10)  \\ \hline
 11 & 1.053(7)  \\ \hline
 12 & 1.044(5)  \\ \hline
 16 & 1.019(2)  \\ \hline
 20 & 1.008(3)  \\ \hline
 24 & 1.003(3)  \\ \hline
\end{tabular}
\caption{Results of the linear fits of the first two terms of eq.(\ref{defPhi4}) for $\beta=0.751800$.}
 \label{ztab3a}  
  \end{table}

  \begin{table}[h!]
  \centering
\begin{tabular}{ | c | c | }
\hline 
 $N_{t}$ & $A(N_t)$  \\ \hline
 13 & 1.168(26)  \\ \hline
 14 & 1.127(28)  \\ \hline
 15 & 1.096(8)  \\ \hline
 16 & 1.039(23)  \\ \hline
 18 & 1.034(33)  \\ \hline
 20 & 1.019(18) \\ \hline
 24 & 1.034(18)  \\ \hline
\end{tabular}
\caption{Same as tab.\ref{ztab3a} but for $\beta=0.756427$.}
 \label{ztab3b}  
  \end{table}

  \begin{table}[h!]
  \centering
\begin{tabular}{ | c | c | }
\hline 
 $N_{t}$ & $A(N_t)$  \\ \hline
 17 & 1.247(35)  \\ \hline
 18 & 1.134(9)  \\ \hline
 19 & 1.100(26)  \\ \hline
 20 & 1.114(23)  \\ \hline
 21 & 1.103(26)  \\ \hline
 22 & 1.065(12) \\ \hline
 24 & 1.025(20)  \\ \hline
 32 & 1.005(14)  \\ \hline
 48 & 0.999(37)  \\ \hline
\end{tabular}
\caption{Same as tab.\ref{ztab3a} but for $\beta=0.758266$.}
 \label{ztab3c}  
  \end{table}

\begin{table}[ht]
\centering %
\begin{tabular}{| c | c | c | c |} 
\hline
$\beta$ & $\gamma_3$ & $\gamma_5$ & $\chi^2_r$ \\
\hline
 0.751800  &  -0.00057(3) & 0.00000038(3) & 1.11 \\
\hline
 0.756427  &  -0.00046(3) & 0.00000028(3) & 1.54 \\
 \hline
  $0.758266 $   & -0.00049(3) & 0.00000031(3) & 1.09 \\

\hline
\end{tabular}
\caption{Results of the fits of $A(N_t)$ according to eq.(\ref{zfit1}) truncated at the order $O(x^6)$ i.e. $O(N_t^{12})$ .}
\label{ztab4}
\end{table}

Looking at these results we see that there is
a remarkable agreement between the values of $\gamma_3$ and $\gamma_5$ obtained with this method and those obtained in the previous section with the snake algorithm. We also see, as anticipated, that with this method there is a significative decrease of 
the uncertainty on the determination of $\gamma_3$.  We also see that
the values obtained (with both methods) for $\beta= 0.751800$ do not agree within the errors with those obtained with the other two values of $\beta$. This suggests that, within the precision of our analysis,  $\beta=0.751800$ is still slightly outside the scaling window,  while the data for   $\beta=0.756427$  and $\beta=0.758266$ agree between them thus showing a good scaling behaviour.

We quote as our final result,
\eq
\gamma_3=-0.00048(4) \hskip2cm 
\gamma_5=3.0(4)\times 10^{-7}
\en
obtained combining together the values obtained at  $\beta=0.756427$  and $\beta=0.758266$ with the present approach.
These values are compatible within the errors with the value obtained with the snake algorithm at $\beta=0.756427$.

\begin{figure}[!htb]
\centering
\includegraphics[width=0.95\textwidth, clip]{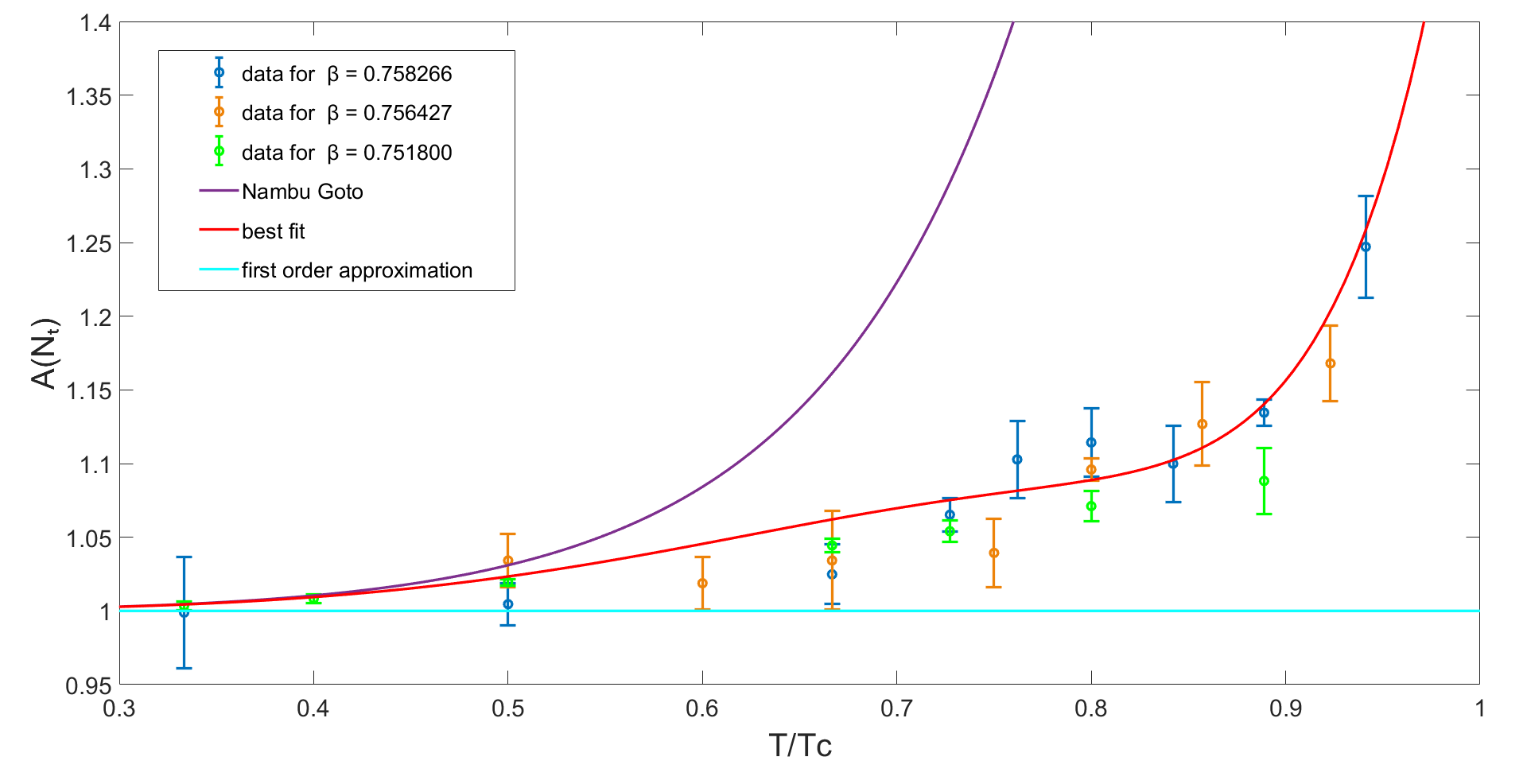}
\caption{Values of $A(N_t)$ for the three different values of $\beta$.  The horizontal line corresponds to the first order approximation 
in which the EST is truncated to the L\"uscher term only,  the violet curve is the Nambu-Goto prediction (truncated at the order $x^6$) and the red one is the prediction of eq.(\ref{zfit1}) with the best fit values of $\gamma_3$ and $\gamma_5$ for $\beta=0.758266 $. }\label{fig_1}
\end{figure}

\section{Concluding remarks}
\label{sec:conclusion}

We studied, using two different methods and different algorithms the correction beyond Nambu-Goto for the confining potential in the three dimensional gauge Ising model. We found a good agreement between the two approaches. The two largest values of $\beta$ that we studied show a good scaling behaviour and  lead to values for $\gamma_3$ and $\gamma_5$ compatible within the errors.  Our final estimate for these parameters is 
$\gamma_3=-0.00048(4)$ and 
$\gamma_5=3.0(4)\times 10^{-7}$.

The value that we obtained for 
$\gamma_3$ agrees with the bound of eq.(\ref{bound1}) , while $\gamma_5$ is slightly below the bound of eq.(\ref{bound2}) which,  inserting the value of $\gamma_3$ and keeping into account the uncertainty in the determination of $\gamma_3$, becomes $\gamma_5>1.6\times 10^{-6}$. This difference is most probably due to the truncation in the perturbative expansion.  Keeping into account higher order terms,  and including also $\gamma_7$ might fill this gap. 

It is interesting to compare our result with the existing estimates for $\gamma_3$ for other LGTs.  In \cite{Caristo:2021tbk} it was shown that also for the $SU(2)$ LGT in three dimension $\gamma_3$ is negative.  The value quoted in \cite{Caristo:2021tbk} is $\gamma_3|_{SU(2)}=- 0.00037(6)$ which is similar, but not compatible within the errors, with the one we obtained here for the Ising model. On the contrary for $SU(6)$ a positive value $\gamma_3 \approx 3\times 10^{-4}$  was found \cite{Athenodorou:2011rx,Dubovsky:2014fma, Chen:2018keo}. These values represent the first steps toward a classification of EST models for LGTs.  Indeed, in the past years, one of the main problems of the EST description of Yang-Mills theories was its universality, i.e. the fact that it predicted essentially the same behavior (with only a mild dependence on the number of spacetime dimensions), for models as different as the three-dimensional $\Z_2$ gauge model and the four-dimensional $\SU(3)$ Yang-Mills theory. This feature is now understood as a universality that manifests itself only at low-energy (or, equivalently, a side-effect of the high accuracy of the Nambu-Goto approximation of EST), while the details related to the gauge group (and, possibly, to the confining mechanism into play) are instead encoded in the $\gamma_i$ corrections, which are not expected to be universal. In particular, from the values quoted above, it seems that $\gamma_3$ for ordinary LGTs takes very small values, which seem to increase with the complexity and size of the gauge group.  This should be contrasted with the case of the 3d $U(1)$ model where sizeable deviations from the Nambu-Goto prediction were observed in several 
quantities \cite{Caselle:2014eka,Caselle:2016mqu,Caselle:2019khe}  which most likely point to a much larger, positive, value of $\gamma_3$.

Finally,  let us add a final comment on the numerical side of our analysis.
As we have seen, duality plays a crucial role in our analysis and for this reason the approach discussed in this paper
is particularly suited for abelian gauge theories, however, apart from the numerical convenience, there is no obstruction to extend the flux based method discussed in sect.\ref{sec:Zago}, given enough computational power, also to non-abelian models. Moreover we have seen that
with this approach the error in the determination of $A(N_t)$ is dominated by the statistical uncertainty while the systematic error due to $\sigma$ and $\alpha$ is fully negligible. This means that there would be in principle no obstruction to improve the estimates of $\gamma_3$ and $\gamma_5$ 
 with a larger statistics.

\newpage

\appendix
\section{Appendix:  Evaluation of $\alpha(\beta)$}
\label{AppendixA}

A first approximation for $\alpha$ can be obtained 
assuming the known scaling behaviour for the string tension in the 3d gauge Ising model 
\eq
\sigma(\beta)=\sigma_c(\beta_c-\beta)^{2\nu}
\nonumber
\en
which leads to a  simple and elegant expression for $\alpha$

\eq
\alpha(\beta)= \frac{2\nu \sigma}{3N_s^2(\beta_c-\beta)}
\nonumber
\label{naive}
\en

However this is not enough for our purposes.
In order to estimate higher order effective string corrections  we need to evaluate 
the flux density with an uncertainty smaller than 1\% and thus 
it is mandatory to include in the expression the next to leading terms of the scaling function.
Following \cite{Caselle:2010pf} and using the results of ~\cite{Caselle:2007yc} we can approximate the scaling function as
\begin{equation}
\label{powerfit}
\sigma(\beta) = \sigma_c t^{2\nu} \times (1 + a t^{\theta} + b t) \;,
\nonumber
\end{equation}
where $t=\tilde\beta-\tilde\beta_{c}$ is the dual of the reduced temperature, $\theta=0.5241(33)$ is the next to leading scaling exponent  and  
the constants take the following values:
 $\sigma_c=10.083(8)$, $a=-0.479(26)$ and $b=-2.12(19)$.
 
Inserting this correction in the definition of $\alpha$ and approximating for simplicity $2\theta\sim 1$ 
 we finally obtain
 \eq
\alpha(\beta)= \frac{2\nu \sigma}{3N_s^2(\beta_c-\beta)}
\left[1+\frac{a\theta}{2\nu}t^\theta+\frac{(b-a^2)\theta}{2\nu}t\right]~~.
\nonumber
\label{subleading}
\en

This is the expression that we used to obtain the values listed in tab.\ref{ztab1}.

\vskip 1.5cm
\noindent {\large {\bf Acknowledgments}}
\vskip 0.2cm
We thank F.Gliozzi, A. Guerrieri, A.Nada and M. Panero for several useful discussions and for a careful reading of the first version of the draft.  
We acknowledge support from the SFT Scientific Initiative of INFN.
This work was partially supported by the Simons Foundation grant
994300 (Simons Collaboration on Confinement and QCD Strings) and by the Prin 2022 grant 2022ZTPK4E .
\vskip 1cm
%


\providecommand{\href}[2]{#2}\begingroup\raggedright\endgroup

\end{document}